# A Phenomenological Relationship Between Vertical Air Motion and Disdrometer Derived *A-b* Coefficients

John Lane [a*], Takis Kasparis [b], Silas Michaelides [b], Philip Metzger [c]

[a] *Ascentech-ESC, Kennedy Space Center Florida, USA*

[b] *Cyprus University of Technology, Limassol, Cyprus*

[c] *Florida Space Institute, University of Central Florida, Orlando, USA*

[*] Corresponding author at: Ascentech-ESC, Kennedy Space Center Florida, USA

*E-mail address*: john.e.lane@nasa.gov (Lane)

ABSTRACT

Using the well-known *Z-R* power law, $Z = A\, R^b$, *A-b* parameters derived from a single disdrometer are readily found and can provide useful information to study rainfall drop size distributions (DSDs). However, large variations in values are often seen when comparing *A-b* sets from various researchers. Values of *b* typically range from 1.25 to 1.55 for both stratiform and convective events. The values of *A* approximately fall into three groups: 150 to 200 for convective, 200 to 400 for stratiform, and 400 to 500 for convective. Computing the *A-b* parameters using the gamma DSD, coupled with a modified drop terminal velocity model, $v_D(D) = v_T(D) - w$, where *D* is drop diameter, $v_T(D)$ is still air drop terminal velocity, and *w* is an estimate of vertical velocity of the air well above the disdrometer, shows an interesting result. This model predicts three regions of *A*, corresponding to $w < 0$, $w = 0$, and $w > 0$. Additional models that incorporate a constant vertical air velocity are also investigated. *A-b* sets derived from a Joss Waldvogel (JW) disdrometer and DSD data acquired near Athalassa, Cyprus, using selected 24-hour data sets from 2011 to 2014, are compared to the above models. The data is separated into two main groups: stratiform events defined by rainfall rates that did not exceed 10 mm h$^{-1}$ at any time during the 24-hour period, and convective events defined by rainfall rates not flagged as stratiform. The convective rainfall is further separated into two groups: *A-b* pairs that fall to the left of the stratiform pairs and pairs that fall to the right. This procedure is repeated with data from other researchers that corresponds to seasonal averages. In all cases, the three vertical groupings of the *A-b* parameter plot seem to correlate to DSD simulations where various values of positive and negative vertical velocities are used.

## 1. Introduction

Research on precipitation advances our ability to understand the components of the water cycle and their respective underlying mechanisms. In this respect, improving precipitation observing methods and systems at both the global and local scale, will ultimately affect the quality of precipitation-related products employed in applications across all scales, with consequential improvements in hydrologic forecasting and water resources management (see Michaelides et al., 2009). Precipitation is characterized by its drop size distribution. All or certainly most precipitation observing systems are based on measuring some aspect of the DSD. Two common classes of observing systems are those that measure the DSD flux, including rain gauges and disdrometers, and those that measure the aerial DSD. Flux measurements can be expressed as a fractional DSD moment, dependent on the choice of raindrop terminal velocity model. If the terminal velocity is proportional to $D^{\gamma}$, rainfall rate $R$ [m s$^{-1}$] is then equal to the $(3+\gamma)^{th}$ moment (see Caracciolo et al., 2006). Weather radar measures reflectivity $Z$ [m$^{-3}$mm$^6$], proportional to the 6$^{th}$ moment of the DSD, while the less common optical extinction $\sigma$ [m$^{-1}$] measures the DSD's 2$^{nd}$ moment (Atlas, 1953; Shipley et al., 1974).

Fundamentally, the drop size distribution dictates the behavior of *Z* and *R*; *Z* and *R* are related through the well-known *Z-R* power law, $Z = A R^b$, that uses two parameters, *A* and *b*. It could be argued that characterizing the *A-b* parameter space is not a useful pursuit since it must be assumed that the *Z-R* relationship is based on a simple two parameter power law. For example, Chapon et al. (2008) studied the *Z-R* relation using a ground based disdrometer and radar data in southern France. The *Z-R* relationships derived from this DSD dataset were found to be very diverse. However, the *A-b* parameter model is an intrinsic part of the US National Weather Service (NWS) radar system and forecasting strategy (Wilson et al., 1979; Choy et al., 1996). In other parts of the world, Ochou et al. (2006) collected Joss-Waldvogel (JW) disdrometer data at 4 sites in western Africa. Using a log-normal distribution model, they derived 4 sets of *A-b* parameters based on long term averages at each of the 4 data collection sites.

It is not just the DSD (usually represented by *N(D)*, where *D* refers to the drop size) that is important in rainfall studies; it is the DSD flux or drop flux distribution (DFD) on a surface that must also be known. In a sense, the *Z-R* relation could be considered a DSD-DFD relation.

The connection between these two quantities is the drop velocity function $v_D(D)$. Note that caution must be exercised in using $v_D(D)$ since it is a vector quantity but is often treated as a scalar function. In Cartesian coordinates, the vector drop velocity $\mathbf{v}_D(D)$ can be decomposed into still air drop terminal velocity $v_T(D)$ in the direction of gravity, vertical component of air motion $w$, and horizontal components of air motion:

$$\mathbf{v}_D(D) = u_x\hat{\mathbf{i}} + u_y\hat{\mathbf{j}} + (w - v_T(D))\hat{\mathbf{k}} \quad , \tag{1}$$

where $u_x$ and $u_y$ are the orthogonal components of the horizontal air velocity $u$. In general, $u$ and $w$ are functions of time and position. The DFD is then the quantity $N(D)v_D(D)$.

A problem with Eq. (1) is that the terminal velocity term $v_T(D)$ is only time independent as shown when $u$ and $w$ are both constant, and enough time has elapsed after the start of a drop trajectory that the sum of the drop's external forces are zero. For the most part, $Z$, $R$, $N(D)$, and $v_D(D)$ are Eulerian quantities since they are measured at fixed points in space and are based on a distribution of particles. The approach taken in this work, is to employ Lagrangian particle trajectory modeling to resolve this problem. This is especially important when modeling the drop velocity at the surface where a rain gauge or disdrometer would be located.

The drop size distribution $N(D)$ has traditionally been modeled as an exponential function or gamma function. The gamma drop size distribution is represented by three parameters, namely, $\mu$, $N_0$, and $\Lambda$ (Ulbrich, 1983) as follows:

$$N(D) = D^\mu N_0 e^{-\Lambda D} \quad . \tag{2}$$

The gamma distribution reduces to the exponential distribution for $\mu = 0$. Numerous researchers have investigated correlations between the DSD shape and observable rainfall characteristics and processes (Rigby et al., 1954; Thurai et al., 2016). Some researchers have investigated relationships between the gamma distribution shape factor and physical processes such as coalescence (Hardy, 1962) and supersaturated updrafts (Igel et al., 2017).

Segregating rain types by means of disdrometer data has long been an active area of research. Using a normalized gamma distribution model, Marzano et al. (2010) investigated the latitude dependence of stratiform and convective rain types, as well as wet and dry periods using a JW disdrometer along with multi-frequency microwave radiometers and microwave polarimetric radar. Islam et al. (2012) using 7 years of JW disdrometer processed by a normalized gamma model, investigated warm, cold, wet, and dry rain types in the southern UK. More recently, Thurai et al. (2016) developed a robust stratiform-convective identification algorithm using two dimensions video disdrometer (2DVD) data, which was tested at sites in Ontario and Huntsville Alabama.

The main focus of this work is to incorporate $w$ into a model of the DSD and DFD in order to observe the predicted effects on $A$ and $b$ and compare to the disdrometer derived $A$ and $b$ coefficients. Several models will be put forward, each with some advantages and disadvantages;

these models are compared to the other. In the end, it will be demonstrated that there is a plausible and moderately predictable connection between the disdrometer derived *A-b* pair and the rainfall type, from which the sign and value of *w* can be estimated.

This paper is structured as follows: In Section 2, an overview of the traditional Marshall-Palmer DSD is given, in order to set the scene for the sections that follow. Section 3 presents a set of three DSD models, built sequentially, where the vertical component of air motion is incorporated into the raindrop terminal velocity and the drop size distribution. The results of the model outputs are compared to disdrometer derived *A-b* parameter pairs. Section 4 discusses the correlation between the simulation and disdrometer data and shows how the physics-based DSD models of Section 3 compare to the empirical gamma DSD model. Section 5 summarizes the overall results and discusses the successes and failures of each of the three models and associated data analysis; recommendations for future work are included in this section too.

## 2. The Marshall-Palmer DSD

The Marshall-Palmer (MP) DSD (Marshall and Palmer, 1948) is a special case of the *exponential* DSD. Because of the specific form of the MP DSD, consistency in computing rainfall rate from the DSD dictates a very specific form of the drop terminal velocity function which is a power-law described by two parameters, $v(D) = v_0 D^\gamma$ m s$^{-1}$ (with *D* in mm). Using this *v(D)*, and starting with *N(D)* in Eq. (2) with $\mu = 0$, the reflectivity *Z* (for S-band and lower frequency bands where non-Rayleigh scattering effects are negligible) and rainfall rate *R* can be computed as follows:

$$Z = \int_0^\infty D^6 N(D)\, dD$$
$$= N_0\, \Gamma(7)\, \Lambda^{-7} \tag{3}$$

$$R = a_R \int_0^\infty v(D)\, D^3 N(D)\, dD$$
$$= a_R N_0 v_0\, \Gamma(4+\gamma)\, \Lambda^{-4-\gamma} \tag{4}$$

where $a_R$ is a units conversion constant such that *R* is represented in standard units of mm h$^{-1}$: $a_R = 0.0036\, \pi/6$ (the $\pi/6$ is due to the drop volume geometry factor). By convention, the *Z-R* relation is also a power-law of the form $Z = AR^b$. A particular choice of *b* is independent of the DSD variables $N_0$ and $\Lambda$, which also leads to the solution of *A*:

$$b = \frac{7}{4+\gamma}, \tag{5}$$

$$A = N_0^{1-b} \, \Gamma(7) \left(a_R v_0 \, \Gamma(4+\gamma)\right)^{-b} . \tag{6}$$

Eqs. (3) through (6) are general results for the exponential DSD with a power-law terminal velocity function. Also, these results are dependent on the limits of integration, which assumes that raindrop diameter extends from 0 to ∞. A refined approach would incorporate a more realistic size range such as 0 to 6 mm. For an impact disdrometer, a size range would be determined by the limits of the instrument detection such as 0.3 mm to 5.5 mm in the case of the Joss disdrometer.

The MP DSD defines the rate variable $\Lambda$ as a function of $R$: $\Lambda(R) = \alpha R^\beta$, where $\alpha = 4.1$ and $\beta = -0.21$, so that $\Lambda$ has units of $mm^{-1}$ and $R$ is expressed in $mm\, h^{-1}$. Substituting this expression for $\Lambda$ in Eq. (4) leads to the following:

$$\gamma = -\frac{1+4\beta}{\beta} = 0.762 , \tag{7}$$

$$v_0 = \frac{\alpha^{-1/\beta}}{a_R N_0 \, \Gamma(-1/\beta)} = 3.254 . \tag{8}$$

Using these values, with the MP value of $N_0 = 8000\ m^{-3} mm^{-1}$, Eqs. (5) and (6) evaluate to $b = 1.47$ and $A = 296$. Note that $b$ is independent of $\Lambda$ and $N_0$, while $A$ is also independent of $\Lambda$ but dependent on $N_0$.

## 3. DSD model development

The $A$-$b$ results above are not the commonly used values associate with the MP DSD model. These values are the result of applying the definition of $Z$ as the 6th moment of the DSD and $R$ as the $3+\gamma$ moment, where the moments are computed as the complete gamma function, i.e., integration limits over $D$ are from 0 to ∞. This exercise is meant only to provide an entrance into the following methodologies, starting with something that is very familiar. The first step is to closely examine the raindrop velocity under conditions of still air and air with a bulk vertical motion.

### 3.1 Raindrop Terminal Velocity

The raindrop velocity is the first and most important modification to be considered when incorporating vertical air motion into the DSD equation. Particle trajectory software written in Fortran 90 was developed by the NASA Granular Mechanics and Regolith Operations (GMRO) Laboratory, Kennedy Space Center between 2010 and 2012. The goal of this software code, namely, Particle Trajectory code with Qshep (PTQ), was to support engineering analysis of future NASA lunar missions by simulating the effects of high speed regolith particle spray, driven by a rocket exhaust plume, on nearby equipment and instruments (Lane et al., 2008). Fig. 1a demonstrates this problem with an illustration of a robotic lander descending near an Apollo site. Fig. 1b shows a similar situation where the Apollo 12 lunar module, *Intrepid*, landed 183 m from Surveyor 3 at Surveyor Crater. In both cases, rocket exhaust gas scours lunar regolith particles from the surface and accelerates them to high speed causing them to impact nearby sensitive equipment (Immer et al., 2011a; Metzger et al., 2008). The software code was developed to simulate the resulting particle velocity as a function of the particle size and rocket exhaust gas properties (temperature, density, and velocity). To calculate the aerodynamic drag force on individual particles, the software assumed the particles are spheres then multiplied the resulting force by an empirically determined shape factor $S_f$ parameter, a scalar value that modifies the drag coefficient $C'_D = S_f C_D$ for non-spherical grains, where $S_f$ = 2 to 3 (Boiko et al., 2005). This software accurately predicted several features of the blowing regolith as recorded in the lunar landing videos looking down from the Apollo Lunar Modules. These measurements included the ejection angle of the dust above the lunar surface, the velocity of the blowing rocks (since the smaller particles were not individually discernible from that height), and the quantity of eroded soil (estimated by analyzing the terrain scouring beneath the Lunar Modules after landing) (Immer et al., 2011b; Metzger et al., 2011). The drag forces from the software also accurately measured the blowing of foam debris in an anomalous launch pad event in the Space Shuttle launch environment (Metzger et al., 2010), and it was used to identify the size and origin of a falling rock that nearly struck a skydiver at high altitude (Metzger, 2014).

In 2016, PTQ-R (here R designates raindrop particle version) was modified to simulate raindrop terminal velocities under conditions of still air and vertical winds. The first step was to define terrestrial conditions for the atmosphere. Trajectories were then generated for various size spherical water drops, falling a distance of 400 m under Earth gravity. This height insured that terminal velocity is achieved for the largest 6 mm drops. Adjustment of the program input parameters were made to match the terminal velocity of small drops of the Gunn and Kinzer (1949) experimental data.

(*a*) (*b*)

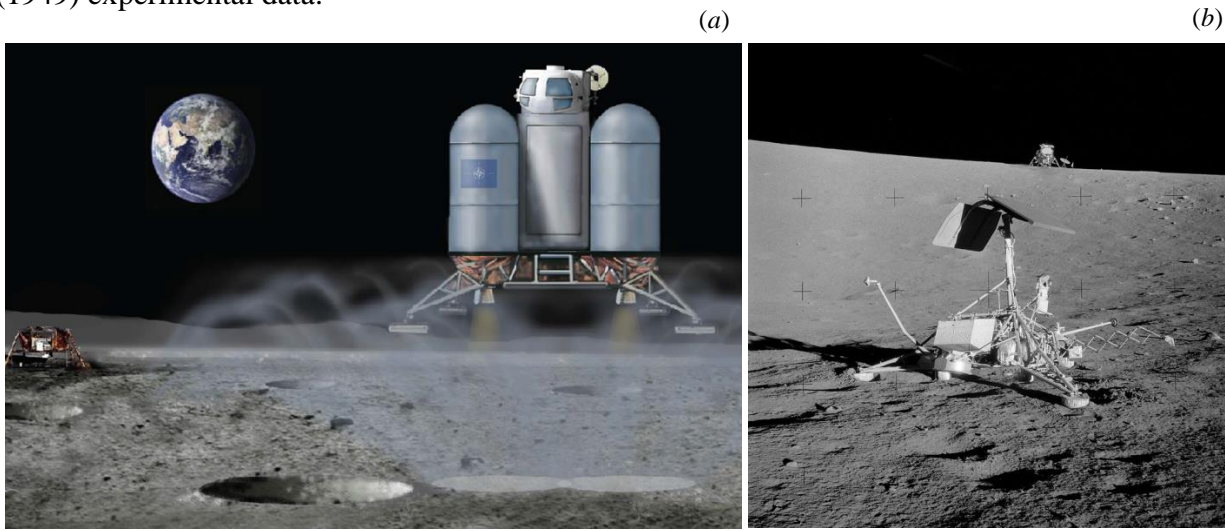

**Fig. 1.** Example application of *Particle Trajectory* code with *Qshep* (PTQ): (a) illustration of a future Google X-Prize class lunar lander descending near an Apollo historical site; (b) Apollo 12 and Surveyor 3 at Surveyor Crater, November 1969.

The next step was to adjust the shape factor $S_f$ to account for the fact that raindrops, unlike lunar soil particles, change shape in response to aerodynamic forces. A relationship was empirically determined relating $S_f$ to particle diameter $D$:

$$S_f = c_0 + c_1 D + c_2 D^2 + c_3 D^3 + c_4 D^4 \qquad , \qquad (9)$$

where $c_0 = 1$, $c_1 = -0.02$, $c_2 = 0.031$, $c_3 = 0$, and $c_4 = 0$, for PTQ-R and $c_0 = 0.665$, $c_1 = 0.956$, $c_2 = -0.564$, $c_3 = 0.125$, and $c_4 = -0.00878$, for PTS-R. Fig. 2a shows the vertical cross-section of a raindrop due to the balance of aerodynamic forces and surface tension on the drop (NASA, 2016). This shape change is directly correlated to the empirical shape factor used to modify the trajectory code drag coefficient. Fig. 2b shows a plot of Eq. (9), corresponding to PTQ-R (thin line). Also shown in this figure is $S_f$ predicted by PTS-R, a stripped down version of PTQ-R available through the NASA software catalog (NASA, 2017). The PTS-R curve (thick line) was fit to a 4$^{th}$ order polynomial of Eq. (9), with a limit that sets the minimum $S_f$ equal to 1. Also shown in Fig. 2b is the drop axis ratio (dotted line) from a 2DVD (Marzuki et al., 2013), where $b$ is the horizontal radius of the drop and $a$ is the vertical radius. More simulation time would be required to refine the raindrop $S_f$ curve, but for the purposes of this work, it will be assumed that difference between the PTQ-R and PTS-R curves represent the scatter in simulation results. Since PTS-R is available through the NASA Software Catalog, its calculations will be used in this paper. The primary difference between PTQ-R and PTS-R is QShep2D interpolation algorithm (Renka, 1988), which due to ACM (Association for Computing Machinery) licensing restrictions was not included in the NASA software release.

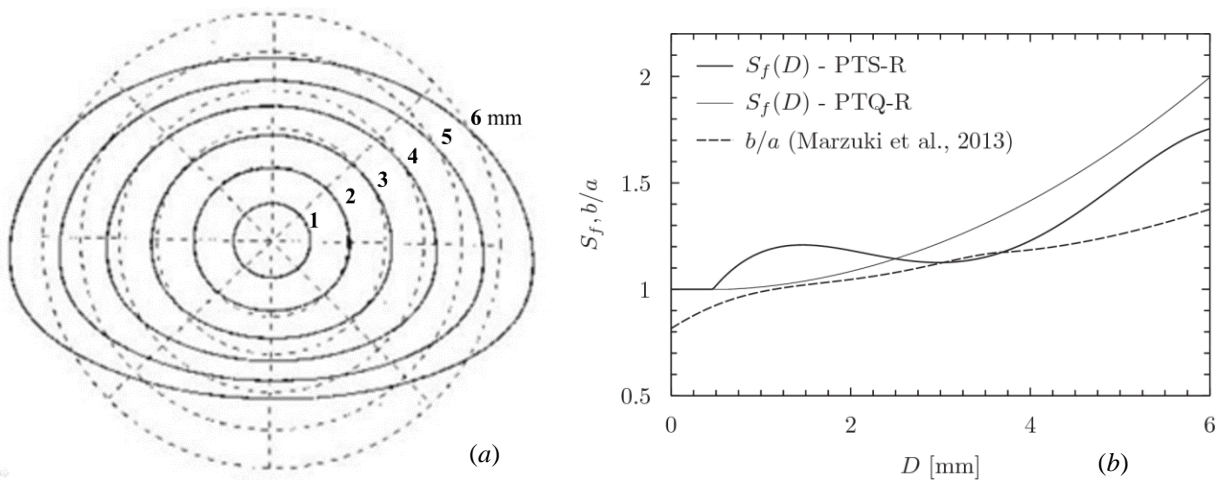

**Fig. 2**. Raindrop shape factor $S_f$ as a function of drop diameter $D$: (a) drop vertical cross-section as a function of $D$; (b) $S_f(D)$ as described by Eq. (9), along with drop axis ratio from a 2DVD.

The output of PTS-R using the shape factor shown in Fig. 2b is plotted in Fig. 3b. Fig. 3a is the input needed to specify the gas (air) properties. For this example, only the vertical velocity component of the gas was not held constant. The density and temperature were set to standard values. The horizontal velocity component of the gas was set to a very small drift value for testing purposes. Fig. 3b shows the PTS-R terminal velocity results for $w = 0$ (black line), $w = -5$ m s$^{-1}$ (downdraft – green line), and $w = +4$ m s$^{-1}$ (updraft – red line), compared to the Gunn and Kinzer (G&K) data. The PTS-R (or PTQ-R) drop velocities at the surface are described reasonably well by the following empirical fit function:

$$v_D(D) = v_1 D + v_2 D^2 + v_3 D^3 - w_0 \left(1 - e^{-D/\lambda}\right) \quad , \tag{10}$$

where $v_1 = 4.67$, $v_2 = -0.789$, and $v_3 = 0.0441$ (for $v_D$ expressed in m s$^{-1}$ and $D$ in mm) fits the G&K data for $w = 0$. The parameter $w_0$ is the vertical wind at an altitude so as not to be influenced by the surface boundary condition where $w(z = 0) = 0$. The terminal velocity of raindrops is modeled as described in Section 3.1. A guess for $w(z)$, the vertical wind profile, is required (see Fig. 3a). Then the graph of Fig. 3b is produced using the PTS-R code. Lastly, a fit is done to the curves of Fig. 3b using Eq. (10), where $\lambda$ is the fitting parameter. There is some (presently unknown) relationship between the $w(z)$ and $\lambda$. For this work, $\lambda$ is treated as a free parameter that is qualitatively linked to the $w(z)$ profile. Future modeling work should be able to quantify this relationship, as well as establish a relationship for the dependencies on air density at altitude $z$.

For this example, $\lambda = 100$ mm generates the curves of Fig. 3b. This is an important result in the case of DSD Model I of the next section - the raindrop terminal velocities at the surface are affected by vertical winds in spite of the $w(z = 0) = 0$ boundary condition because of the finite distance needed to adapt to still air terminal velocity, as the surface boundary condition is approached during drop fall. Note that vertical wind profile $w(z)$ of Fig. 3a, may in general be a complete unknown without some collocated physical measurement, such as a wind profiler, radiosonde, or tower mounted 3-axis anemometers. Since the vertical wind profile is an input of PTS-R, the output of Fig. 3b cannot be generated without it.

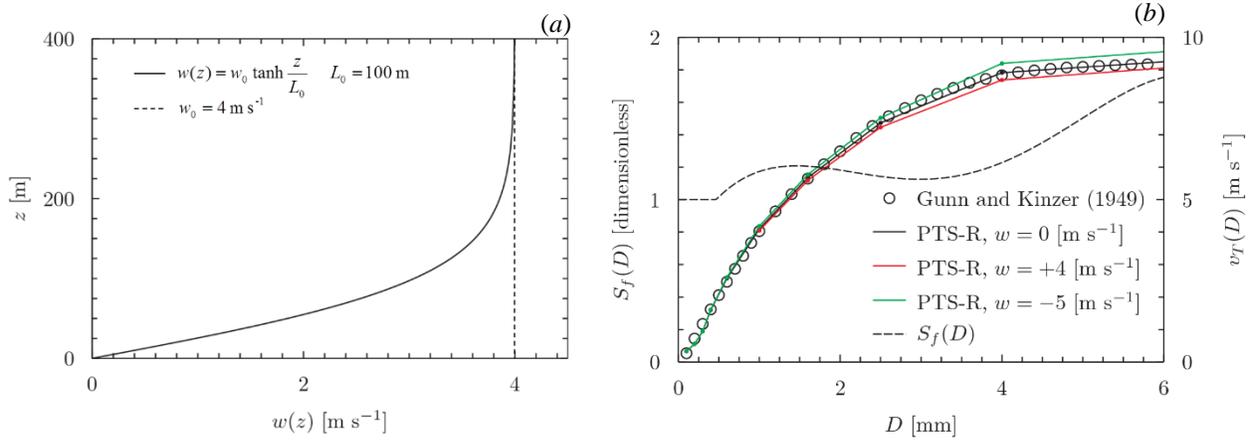

**Fig. 3**. Raindrop terminal velocity based on PTS-R: (a) PTS-R example input of simulated vertical wind profile (updraft); (b) PTS-R terminal velocity results for $w = -5$ m s$^{-1}$ (downdraft), $w = +4$ m s$^{-1}$ (updraft), compared to the Gunn and Kinzer (1949) data.

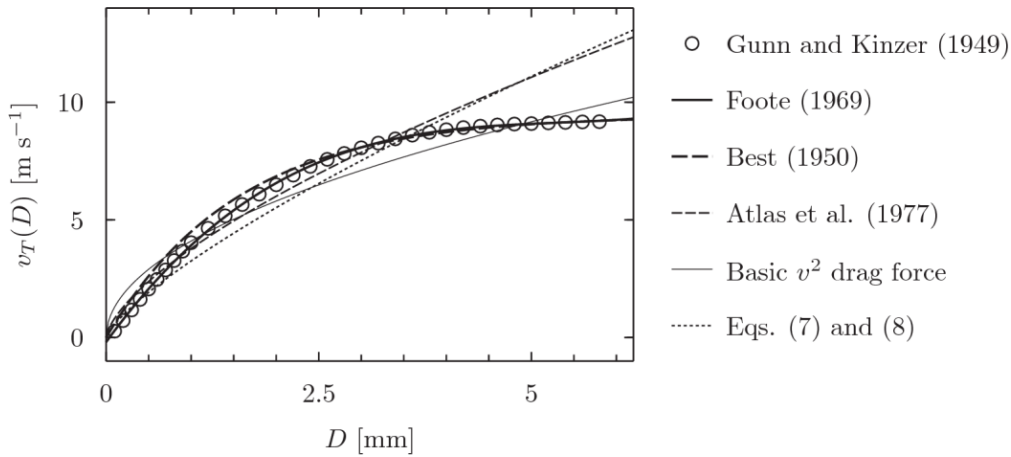

**Fig. 4**. Comparison of other raindrop terminal velocity models with the G&K data.

Fig. 4 displays other popular terminal velocity fitting functions. The fit that most closely matches the data of G&K is the one by Foote and Toit (1969). The next best fit in this group is possibly Best (1950). The other three fitting functions are two-parameter power-laws. Power-laws, such as the one proposed by Atlas and Ulbrich (1977), where $v(D) = 3.78\, D^{2/3}$ m s$^{-1}$, have the advantage of being very easy to work with, but at the expense of inaccuracies. A more accurate formulae, previously proposed by Atlas et al. (1973) is $v(D) = 9.65 - 1030\, e^{-0.6D}$ m s$^{-1}$. Note that the Foote and Toit (1969) function is very similar to Eq. (10) with $w_0 = 0$. Both the Atlas et al. (1973) formulae and the Foote and Toit (1969) polynomial have traded accuracy in the disdrometer detection range with failure to match properly at $D = 0$. Equation (10) is good at $D = 0$ where $v(D) = 0$ with some minimal sacrifice of matching G&K for $D > 0$.

*3.2. Model I*

This approach yields exact equations for $A$ and $b$, analogous to Eqs. (5) and (6) of the MP DSD shown previously. The starting point is the gamma DSD of Eq. (2) and a modified drop velocity function based on a power-law:

$$v(D) = v_0 D^\gamma - w \quad , \tag{11}$$

where $w$ is the speed of the vertical air motion in the volume above the disdrometer site. A positive value of $w$ corresponds to upward air motion, i.e., updraft. A negative value of $w$ is associated with a downdraft. In this model, $w$ is assumed to be a constant and independent of drop size. Since vertical air motion must be zero at the surface, Eq. (11) describes a situation which is only valid above some height $L$ above the ground. Therefore, using this form of terminal velocity for modeling a disdrometer derived DSD will only be strictly true if the disdrometer is mounted on a tower at height $L$ or higher. At the surface, $w$ can be exchanged for a function of $D$ such that $w_D(D) \to 0$ as $D \to 0$. If $L$ is sufficiently small, corresponding to a wind profile where $w(z=L) \approx w(\infty)$, then the value of $w_D(D)$ will be non-zero for larger drop sizes. This effect is depicted in Fig. 3b as the differences between the $w = 0$ case and the red and green lines. This model of $w_D(D)$ is part of Model II that will be described in the next section. For Model I, Eq. (11) will be used with a constant $w$, even though it is clear that the results must be appropriately interpreted.

The advantage of using Eqs. (2) and (11) is that exact algebraic equations for $A$ and $b$ are generated. Similar to that above for the MP DSD, the quantities $Z$ and $R$ are computed as the $6^{th}$ and $3+\gamma$ moments of the DSD, respectively. Note that limits of integration now take into account the observable drop size range, where $D_1$ is a function of $w$:

$$\begin{aligned} Z &= \int_{D_1(w)}^{D_2} D^6 N(D)\, dD \\ &= N_0\, \Gamma(7+\mu, \Lambda D_1(w), \Lambda D_2)\, \Lambda^{-7-\mu} \\ &= N_0\, \Gamma_I(7+\mu)\, \Lambda^{-7-\mu} \end{aligned} \tag{12}$$

The gamma function in Eq. (12) is an *incomplete gamma function* since the limits are no longer from 0 to ∞. For ease of notation, $\Gamma_I(n)$ will be used to represent the incomplete gamma function, $\Gamma(n, x_1, x_2)$, where it will always be assumed unless otherwise noted that $x_1 = \Lambda D_1$ and $x_2 = \Lambda D_2$. Now, $R$ can be computed as follows:

$$R = a_R \int_{D_1(w)}^{D_2} v(D) D^3 N(D) \, dD$$

$$= a_R N_0 \Lambda^{-4-\mu} \left( v_0 \Lambda^{-\gamma} \Gamma_I(4+\mu+\gamma) - w \Gamma_I(4+\mu) \right)$$

(13)

The upper limit of integration $D_2$ is just the maximum drop size that can be detected by the system. In this work, we assume $D_2 = 5.5$ mm. The lower limit $D_1(w)$ can be determined by setting Eq. (11) to 0 and solving for $D_1$:

$$D_1(w) = \begin{cases} D_L & w \le v_0 D_L^{\gamma} \\ (w/v_0)^{1/\gamma} & \text{otherwise} \end{cases}, \quad (14)$$

where $D_L$ is the lower limit of the disdrometer detection ($D_L = 0.3$ mm in this work).

The next step is to choose a formulation for $b$ that is independent of the rain rate parameter $\Lambda$ and $N_0$:

$$b = \frac{7+\mu}{4+\mu+\gamma} \quad . \quad (15)$$

Eq. (15) reduces to Eq. (5), the exponential DSD case, for $\mu = 0$ and the special MP DSD case when $\gamma = 0.762$.

Solving for $A$ is a simple matter of rearranging the $Z$-$R$ power law:

$$A = \frac{Z}{R^b}, \quad (16)$$

where Eqs. (12), (13), and (15) are used explicitly, with Eq. (14) used implicitly. Fig. 5 shows plots of Eqs. (15) and (16) for various values of $N_0$, $\mu$, and $\Lambda$ and three values of vertical velocity: $w = 0$, $w = -4$ m s$^{-1}$, and $w = 2$ m s$^{-1}$. For the range of DSD variables shown, there is a distinct separation of $A$-$b$ positions due to the sign and magnitude of $w$. The relative amount of separation in $A$ due to $w$, is greater with increasing value of $b$.

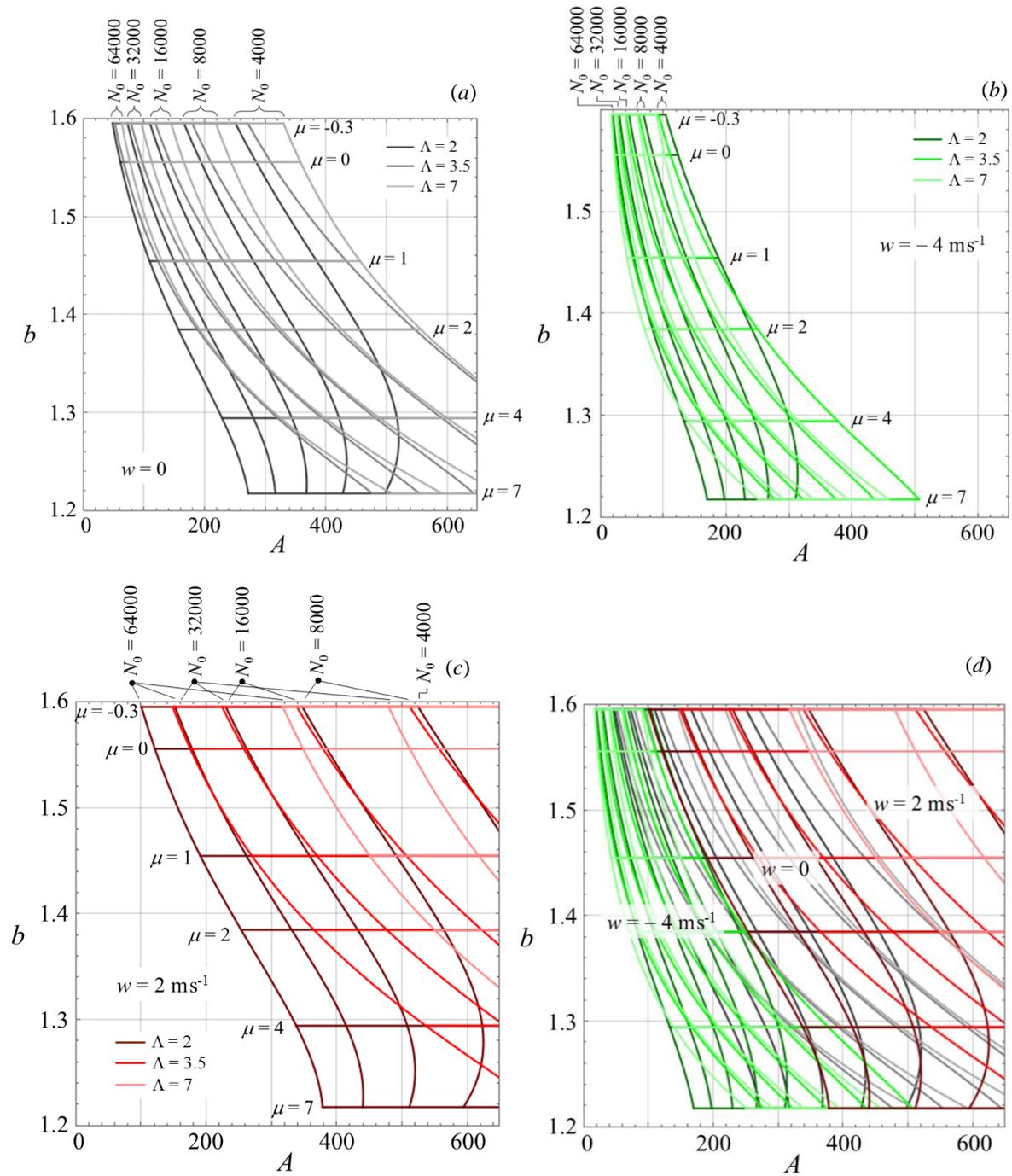

**Fig. 5**. Eqs. (15) and (16) plotted for various values of $N_0$, $\mu$, and $\Lambda$ with, (a) $w = 0$; (b) $w = -4$ m s$^{-1}$; (c) $w = 2$ m s$^{-1}$; (d) composite of all three cases.

Fig. 6 is a composite of Fig. 5d overlaid with $A$-$b$ derived results from disdrometer data. The grey symbols represent stratiform rain, as defined by rainfall rate magnitudes. The red and green symbols represent convective rain, which is everything that is not stratiform. The red symbols separate the green symbols defined by a line at $A = 300$. Fig. 6a is disdrometer derived $A$-$b$ data from Athalassa, Cyprus, selected from events between 2011 and 2014 (Lane et al., 2016). Fig. 6b shows disdrometer derived $A$-$b$ data compiled by Sulochana et al. (2016).

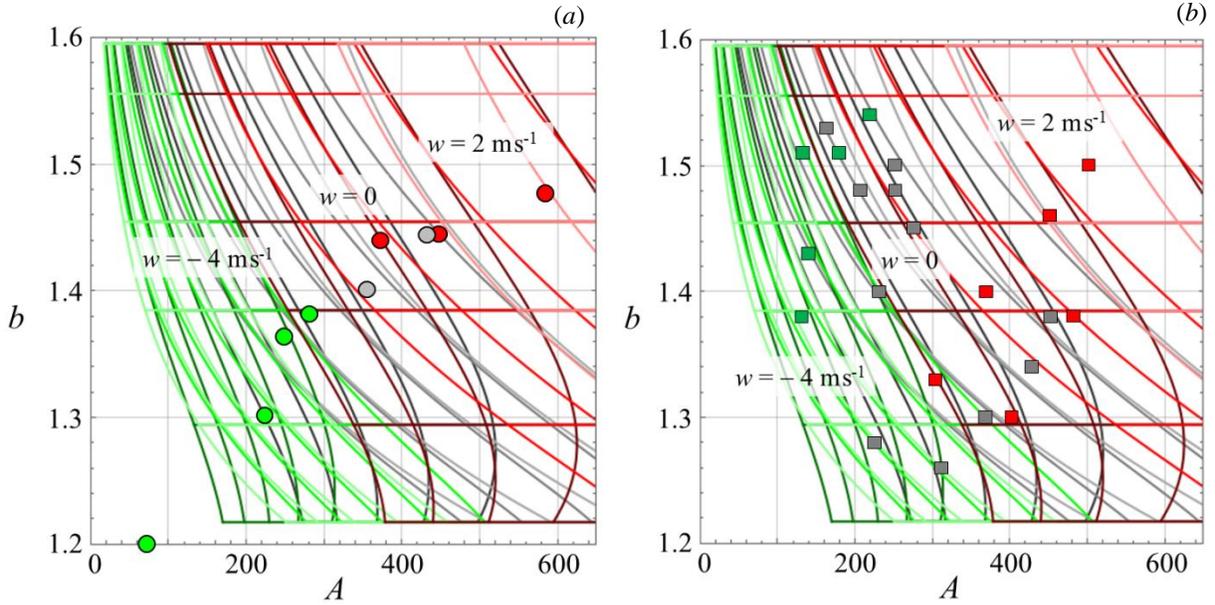

**Fig. 6**. Composite of Fig. 5d overlaid with $A$-$b$ derived results from disdrometer data with grey symbols representing stratiform rain, red symbols for convective rain with $A > 300$, and green symbols for convective rain with $A < 300$: (a) data from Athalassa, Cyprus; (b) data from Sulochana et al. (2016).

*3.3 Model II*

Model II attempts to mitigate some of the problems of the previous model, in particular the boundary condition failure for $w$ at the surface. To accomplish this, Model II incorporates a *continuity equation* approach that can alternatively be described as *flux conservation* where the flux $F(D) = v(D) N(D)$ (Parvez et al., 2002) through any two parallel planes at heights $z_1$ and $z_2$ parallel to the surface is conserved for all $D$. $F(D)$ is exactly the quantity that a disdrometer tries to measure. This pseudo law can also be expressed as:

$$v(z_1, D)\, N_1(D) = v(z_2, D)\, N_2(D) \quad , \tag{17}$$

where $v(z, D)$ is the drop velocity function at height $z$. Atlas et al. (1974) discusses alternatives for terminal velocity models and their dependence on air density and altitude. For the purposes of this work and a manageable Model II, Eq. (17) is simplified by setting $v(z_1, D) = v_T(D)$, the still air terminal velocity at the disdrometer where $z_1 = 0$; $N_1(D) = N_D(D)$, the DSD measured by the disdrometer; $v(z_2, D) = v_T(D) - w$, where $w$ is the vertical air velocity at $z$; and $N_2(D) = N(D)$, the DSD at altitude $z$ above the surface. Vertical velocity $w$ can be modeled as a constant multiplied by the factor $f(\rho) = (\rho_0/\rho)^{0.4}$ as discussed in Foote and du Toit (1969), where $\rho_0$ is the air density at sea level and $\rho$ is the density aloft. For the purposes of this paper, accounting for air density dependencies will be reserved for future work, which is a reasonably valid approach when the distance between $z_1$ and $z_2$ is small enough to neglect the corresponding differences in density so that $f(\rho_1) \approx f(\rho_2) \approx 1$.

Eq. (17) with the above substitutions describes the flux conservation model (FCM) which is the basis of Model II, and can be rearranged to express $N_D(D)$. However, this leads to some problems. The first problem is equivalent to defining $D_1$ in Model I. This time however, a unit step function or *Heaviside* function $H(\chi)$ will be used to prevent negative values of $N_D(D)$:

$$N_D(D) = \chi H(\chi) N(D) \qquad , \qquad (18)$$

where

$$\chi \equiv \frac{v_T(D) - w}{v_T(D)} \qquad . \qquad (19)$$

There still remains a problem: seldom does a measured disdrometer $N_D(D)$ display a sharp cutoff for updrafts ($w > 0$) at a $D_1$ as defined by $v_T(D_1) = w$. The model needs to account for the regeneration of a new DSD near the surface where $w \to 0$. This additional DSD component is most likely the result of drop collision and breakup below a few hundred meters above the surface. The last part of this model is to set $N(D)$ to a purely exponential function of $D$ so now $N_D(D)$ becomes:

$$N_D(D) = \phi(w, \xi, D) N_0 e^{-\Lambda D} \qquad , \qquad (20)$$

where

$$\phi(w, \xi, D) = \chi H(\chi) + \xi \qquad , \qquad (21)$$

and $\xi$ is an empirical parameter that adds a fraction of the original $N(D)$ back to $N_D(D)$, physically attributed to drop breakup and DSD regeneration near the surface.

Using the DSD as defined in Eq. (20) and the drop velocity of Eq. (10) with $w_0 = w$, $Z$ and $R$ expressed as

$$Z = \int_{D_1}^{D_2} D^6 N_D(D) \, dD \qquad , \qquad (22)$$

$$R = a_R \int_{D_1}^{D_2} v_D(D) D^3 N_D(D) \, dD \qquad . \qquad (23)$$

can be computed using a Monte Carlo approach to generate a set of points representing a single rain event. The Monte Carlo varies $N_0$ and $\Lambda$ for each point in the $Z$-$R$ set. The vertical velocity parameter $w$ is held constant throughout. From this set, an $A$-$b$ pair can be found from a linear fit of log$Z$ versus log$R$. This procedure can be repeated many times to generate many $A$-$b$ point pairs. Note that $D_1$ and $D_2$ in Eqs. (22) and (23) are constant and represent the instrument measurement limits. In this work, $D_1 = 0.3$ mm and $D_2 = 5.5$ mm.

Fig. 7 shows the result of a Monte Carlo generation of 1000 $A$-$b$ pairs for each of the three velocities shown, grey is $w = 0$, red is $w = 2$ m s$^{-1}$, and green is $w = -4$ m s$^{-1}$. These particular values were chosen so a comparison could be made to the results of Model I. Constant parameters were $\lambda = 40$ mm in Eq. (10) and $\xi = 0.2$ in Eq. (21). The circles and squares are the same data shown in Figs. 6a and 6b.

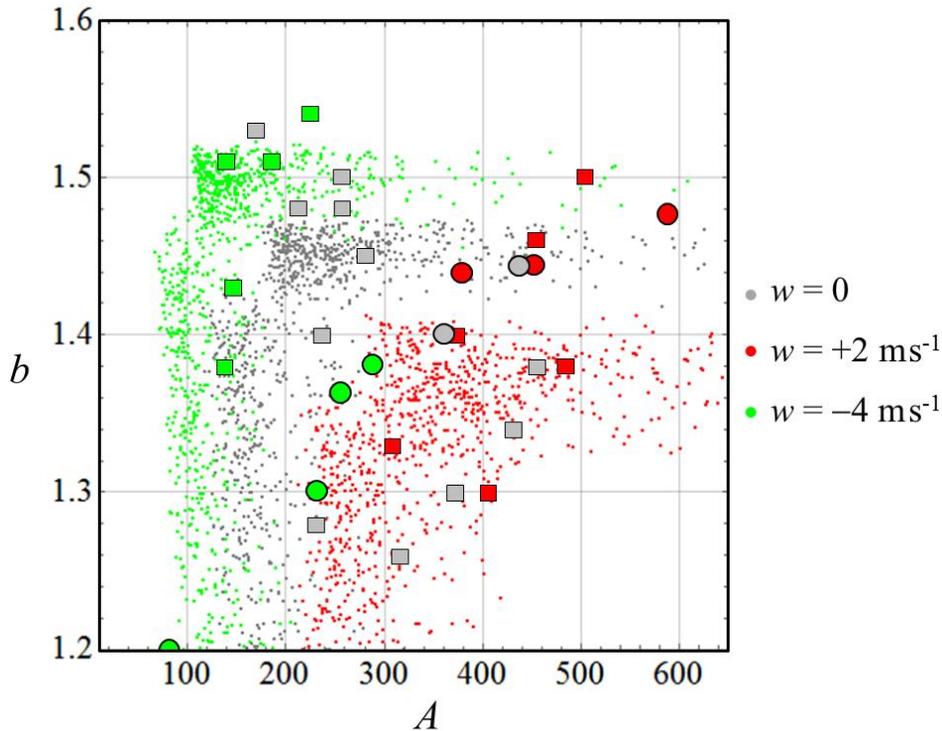

**Fig. 7**. Result of a Monte Carlo generation of 1000 $A$-$b$ pairs for each of the three velocities shown. Constant parameters are $\lambda = 40$ mm in Eq. (10) and $\xi = 0.2$ in Eq. (21).

*3.4 Model III*

The third and final DSD model that will be considered is just a sum of two models described by Eq. (20):

$$N_D(D) = \phi(w_1, \xi_1, D) N_1 e^{-\Lambda_1 D} + \phi(w_2, \xi_2, D) N_2 e^{-\Lambda_2 D} \quad . \tag{24}$$

The rationale behind Eq. (24) is that both updrafts and downdrafts might be experienced by a measurement site during a single storm event. The disdrometer data from Athalassa,

Cyprus was processed in 24 h sets, which may include multiple rain events, so it is therefore logical to expect multiple DSD characteristics and/or vertical wind magnitudes and directions for each data set. Multimodal DSDs and vertical wind effects have been investigated by other researchers (see for example, Ekerete et al., 2016).

Figure 8 shows 11 selected sets of 24 h data from the JW disdrometer located in Athalassa, Cyprus [for a brief description of the principles of the JW disdrometer operation, the reader may refer to Michaelides et al. (2009)]. The event tag name is the disdrometer data date in day-month-year format (e.g., the event tag 050714 represents 5 July 2014). The left graph in each set is $N(D)$ computed from the disdrometer data using Eq. (10) with $w_0 = 0$. The red solid line is a fit of Eq. (24). The right graph is the disdrometer derived rainfall rate at 10 s and 60 s intervals. Table 1 is a summary of the fit parameters in Fig. 8. The last column is the percentage of rainfall accumulation from the two independent terms, $DSD_1$ and $DSD_2$ of Eq. (24). In all cases, $DSD_1$ is the primary contributor to rainfall accumulation.

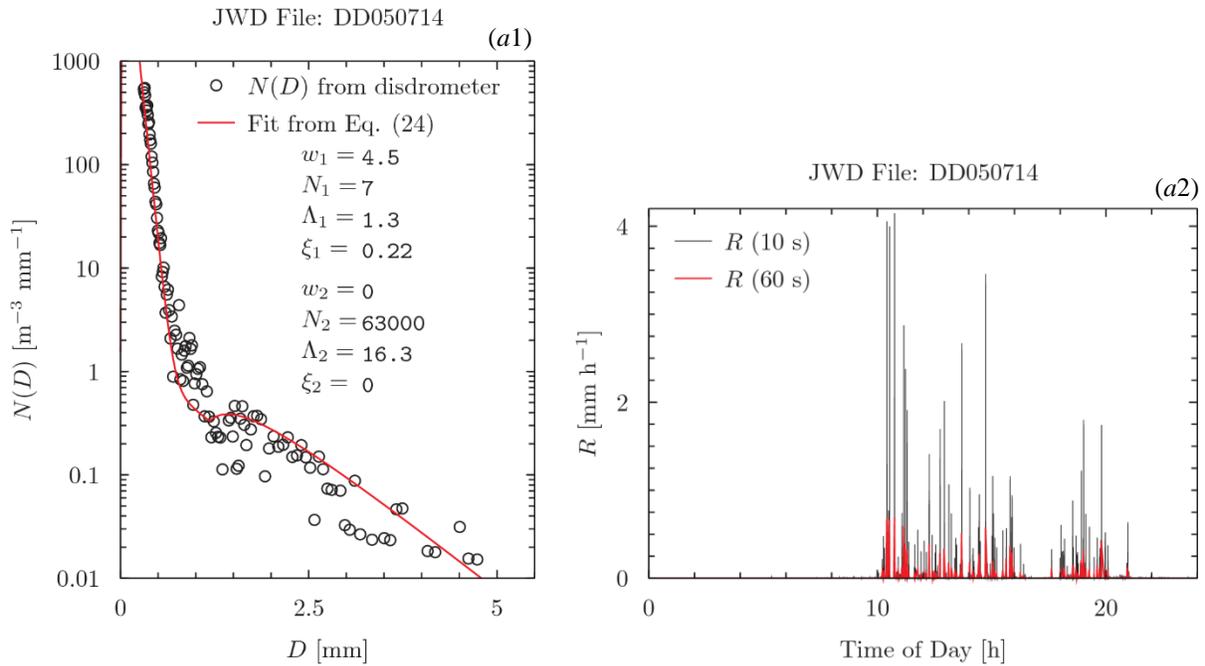

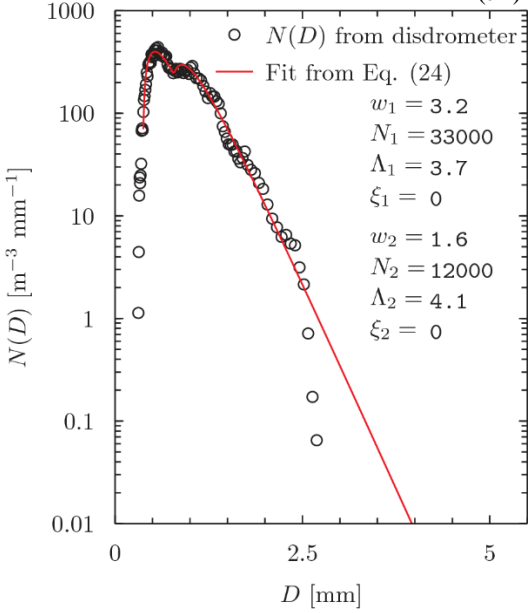

(b1)

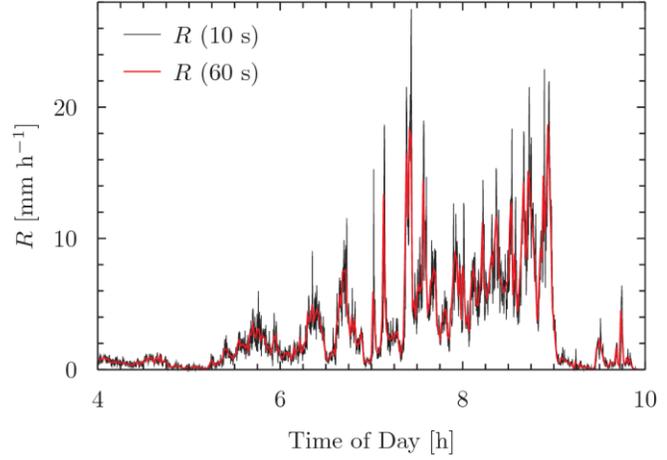

(b2)

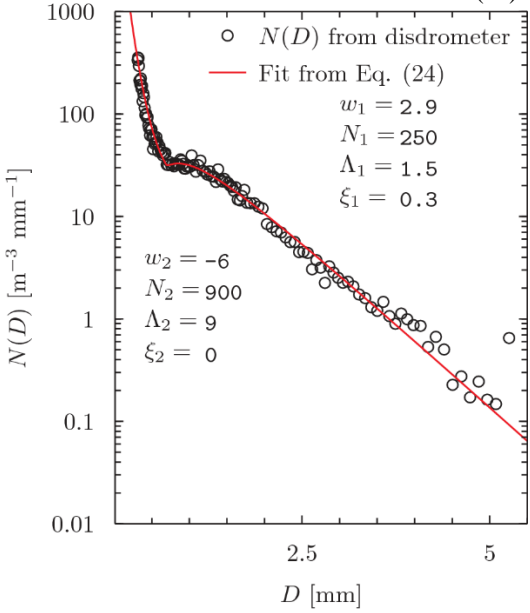

(c1)

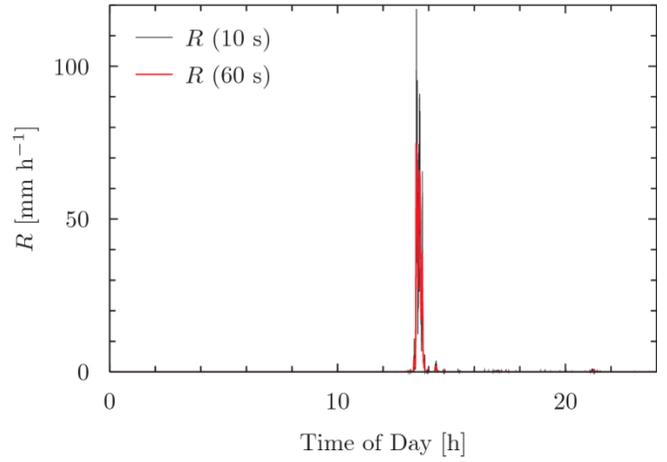

(c2)

(d1)

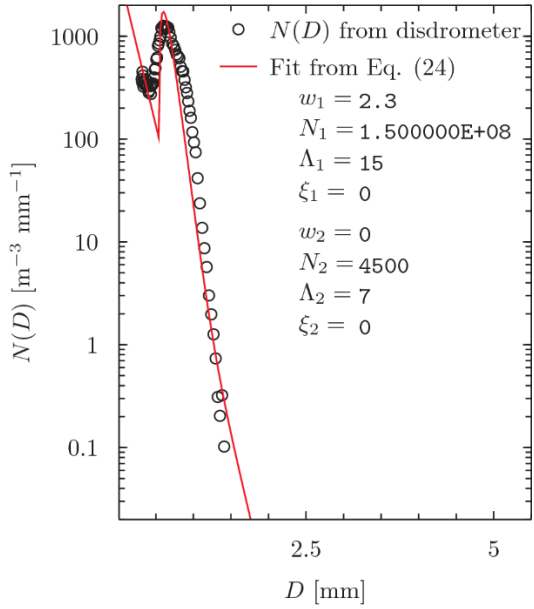
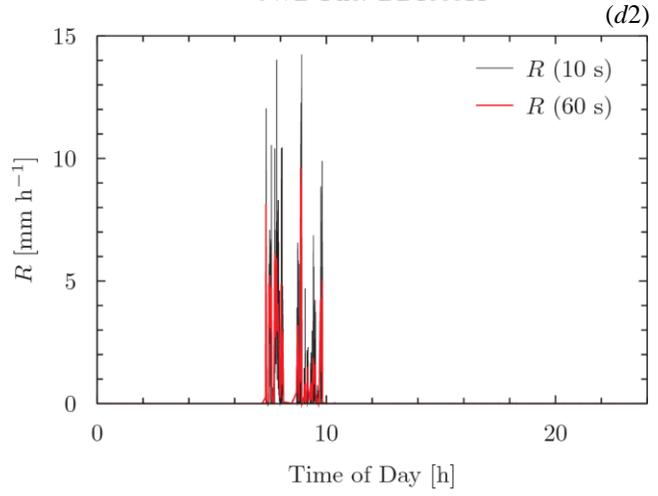
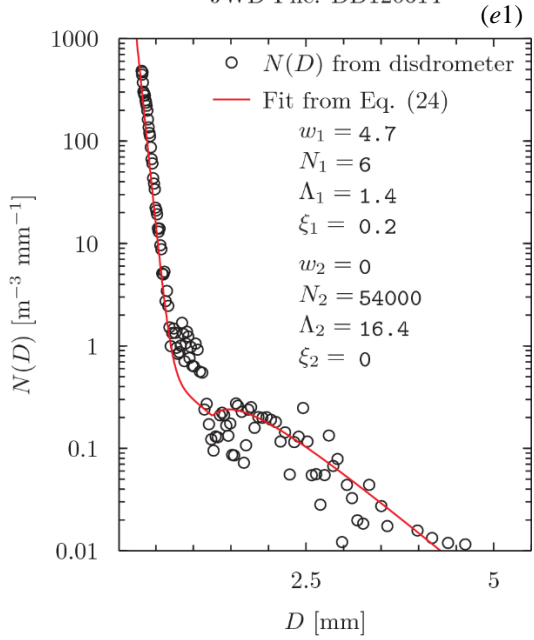
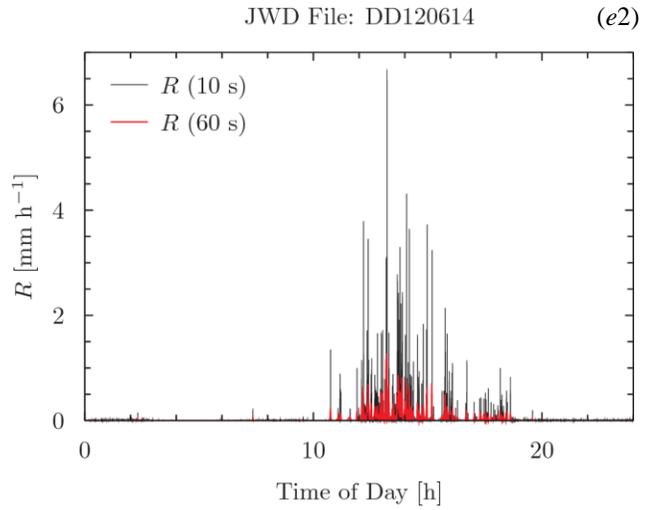

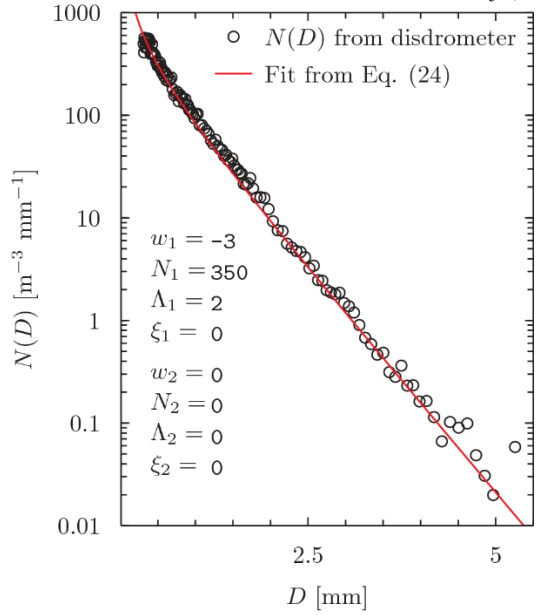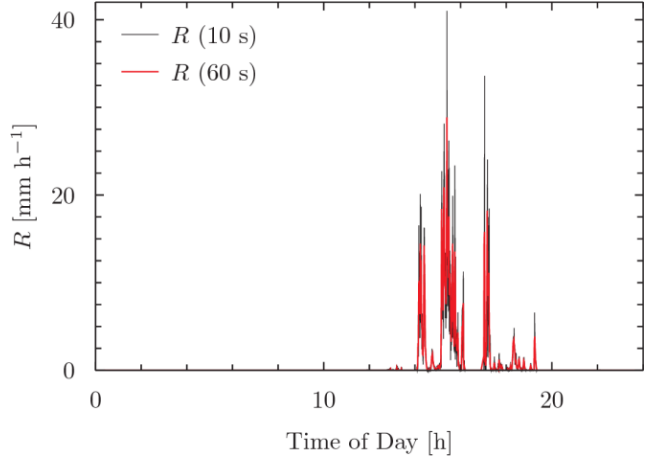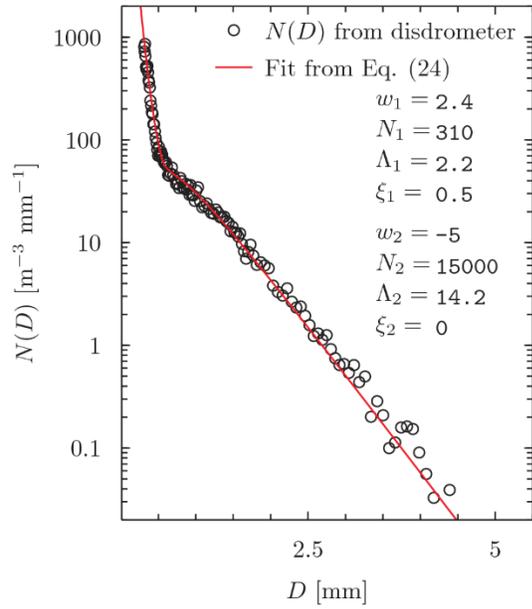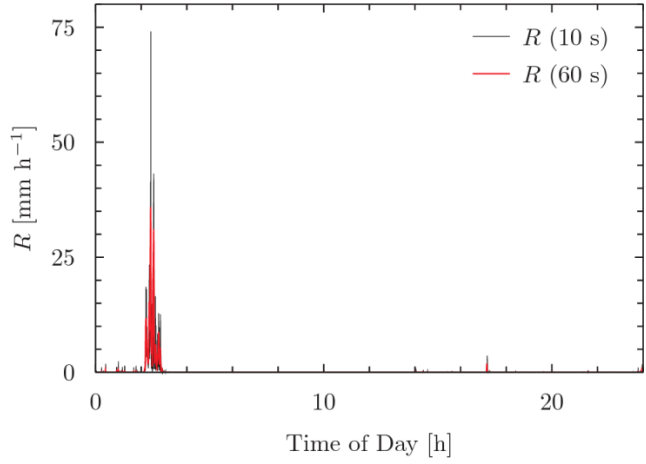

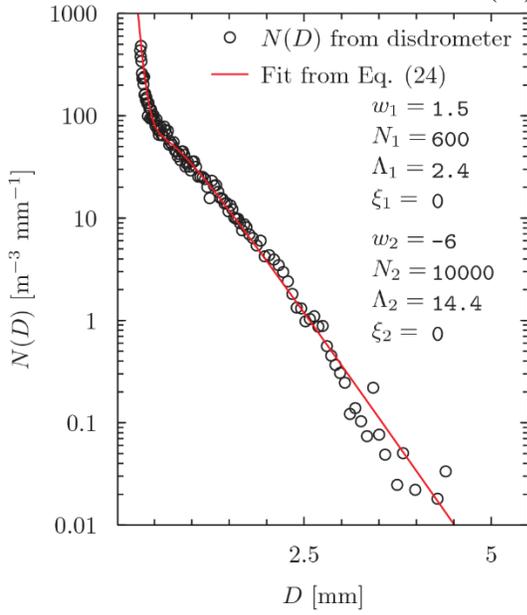
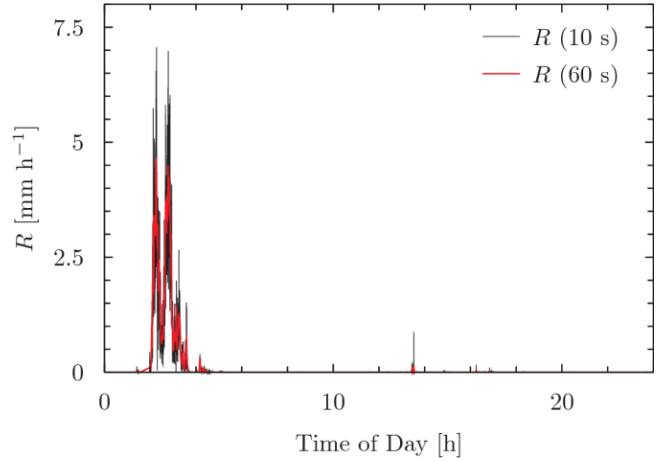
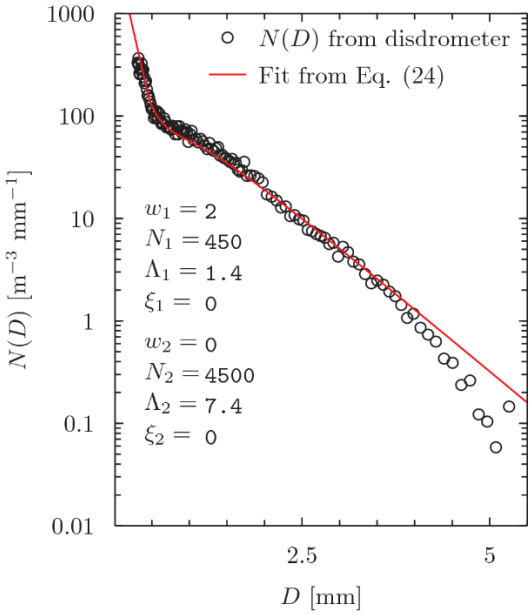
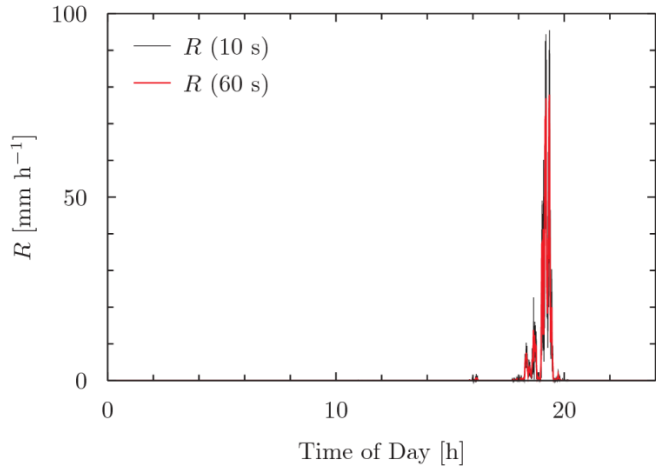

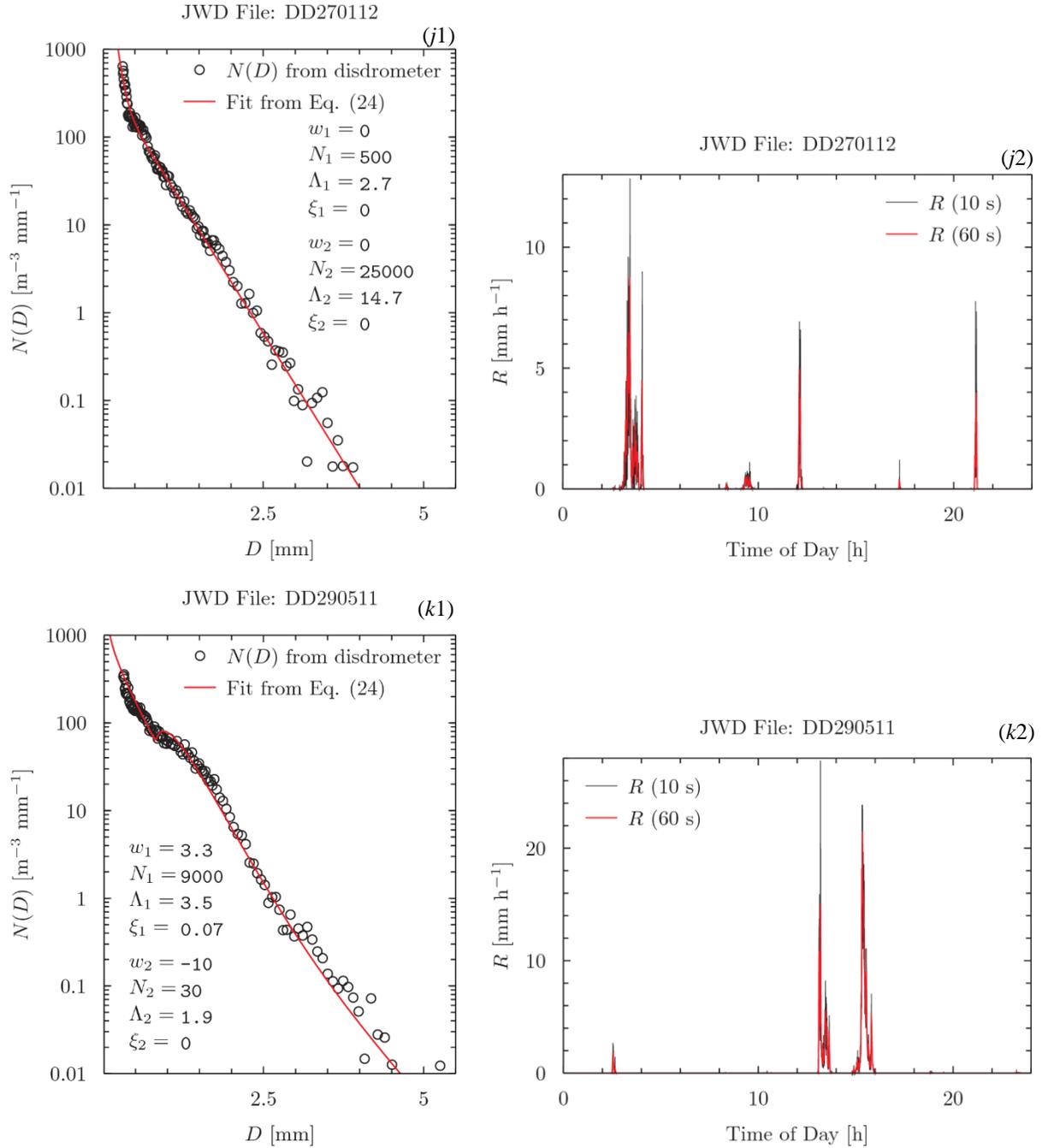

**Fig 8**. Model III fit of Athalassa data, as described by Eq. (24).

**Table 1**

Summary of fit parameters of the data in Fig. 8.

| Event Tag ($\Delta t$ = 24 h) | $w_1 \setminus w_2$ m s$^{-1}$ | $N_1 \setminus N_2$ m$^{-3}$ mm$^{-1}$ | $\Lambda_1 \setminus \Lambda_2$ mm$^{-1}$ | $\xi_1 \setminus \xi_2$ unitless | $A$ | $b$ | $R_{max}$ (10 s $\setminus$ 60 s) mm h$^{-1}$ | $RA_1 \setminus RA_2$ % |
|---|---|---|---|---|---|---|---|---|
| DD050714 | 4.5 \ 0 | 7 \ 63000 | 1.3 \ 16.3 | 0.22 \ 0 | 1765 | 1.63 | 4.1 \ 0.7 | 99.7 \ 0.3 |
| DD060613 | 3.2 \ 1.6 | 33000 \ 12000 | 3.7 \ 4.1 | 0 \ 0 | 221 | 1.30 | 28 \ 18 | 68.9 \ 31.1 |
| DD100513 | 2.9 \ −6 | 250 \ 900 | 1.5 \ 9.0 | 0.3 \ 0 | 584 | 1.47 | 120 \ 75 | 99.9 \ 0.1 |
| DD100811 | 2.3 \ 0 | $1.5 \times 10^8$ \ 4500 | 15 \ 7 | 0 \ 0 | 71 | 1.20 | 15 \ 9.7 | 93.4 \ 6.6 |
| DD120614 | 4.7 \ 0 | 6 \ 54000 | 1.4 \ 16.4 | 0.2 \ 0 | 1391 | 1.64 | 4.3 \ 1.3 | 99.5 \ 0.5 |
| DD170413 | −3.0 \ 0 | 350 \ 0 | 2 \ - | 0 \ - | 280 | 1.38 | 42 \ 28 | 100 \ 0 |
| DD181013 | 2.4 \ −5 | 310 \ 15000 | 2.2 \ 14.2 | 0.5 \ 0 | 373 | 1.42 | 74 \ 35 | 99.9 \ 0.1 |
| DD230911 | 1.5 \ −5 | 600 \ 10000 | 2.4 \ 14.4 | 0 \ 0 | 432 | 1.43 | 7.0 \ 4.7 | 99.9 \ 0.1 |
| DD241012 | 2.0 \ 0 | 450 \ 4500 | 1.4 \ 7.4 | 0 \ 0 | 446 | 1.43 | 95 \ 77 | 99.7 \ 0.3 |
| DD270112 | 0 \ 0 | 500 \ 25000 | 2.7 \ 14.7 | 0 \ 0 | 354 | 1.40 | 12.9 \ 8.8 | 99.9 \ 0.1 |
| DD290511 | 3.3 \ −10 | 9000 \ 30 | 3.5 \ 1.9 | 0.07 \ 0 | 248 | 1.36 | 24 \ 22 | 70.2 \ 29.8 |

## 4. Discussion

Fig. 9 summarizes the results using Model III as a fit to the 11 selected events which were recorded at Athalassa, Cyprus (Latitude 34º55'N (34.92ºN) and Longitude 32º20'E (32.33ºE). The color coding in this figure follows a different rule. Whereas Figs. 6a and 6b were based on the *A* value greater than or less than 300, the color scheme here is based on the vertical wind direction from the fit reported in Table 1. As before, grey indicates rainfall rate less than 10 mm h$^{-1}$. However, if *w* is greater than 1.5 m s$^{−1}$ then the symbol will be shown as a dual color. For example the points in the upper right and lower left, have low rainfall rate < 10 mm but also have high positive *w* values. Therefore, these are color coded as grey and red. One of the points, corresponding to event DD290511, has positive and negative value of *w*, corresponding to DSD$_1$ and DSD$_2$ respectively, is therefore color coded red and green. The open symbols are other *A-b* pairs. The open square is the standard NWS *summer deep convection Z-R* relation, *A* = 300 and *b* = 1.4. The open triangle is the standard *Marshall-Palmer stratiform Z-R* relation, *A* = 200 and *b* = 1.6 (see *Z-R* Relationship Tables, National Weather Service (2016)). The open circle is the *A-b* pair derived above using the MP DSD, from Eqs. (5) and (6).

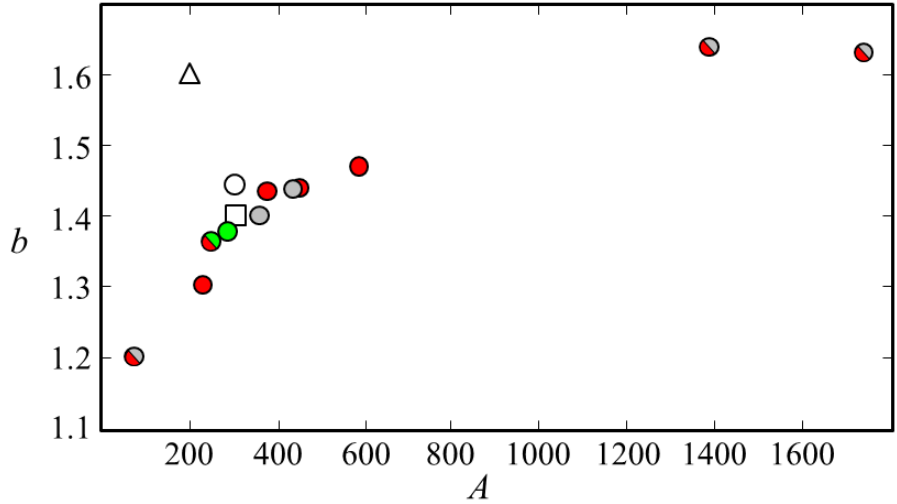

**Fig. 9**. Results of Model III, based on the 11 events in Table 1. Bi-colored symbols indicate events with dual characteristics. Open symbols are other *A-b* pairs: square is NWS summer deep convection; triangle is the Marshall-Palmer stratiform; circle is *A-b* derived above using the MP DSD.

Rigby et al. (1954) investigated effects of drop coalescence, accretion, and evaporation versus DSD shape. Hardy (1963) conducted similar investigations and concluded that the number of smaller drops is depleted by each of these processes; the number of larger drops is increased by coalescence and accretion but decreased by evaporation. When the slope of the DSD is small, the effect of these processes is diminished. The results of both Rigby and Hardy can now be correlated to the shape parameter $\mu$ of the gamma DSD (however the gamma DSD was not being considered at the time of those writings). In this current work, and for simplicity, only one mechanism is under investigation, where the sign and magnitude of $\mu$ is correlated to the sign and magnitude of *w*.

The gamma DSD model of Eq. (2) and the flux conservation DSD model of Eq. (20), both share the same exponential function but the prefactors of the exponential at first appear to be very different. Eq. (25) defines these two cases:

$$f(D) \equiv D^{\mu} \quad \text{and} \quad g(D) \equiv \chi H(\chi) + \xi \quad , \qquad (25)$$

where *f(D)* is the gamma DSD prefactor and *g(D)* is the flux conservation DSD prefactor. The function *g(D)* is dependent on *w* as well as the parameter $\xi$. The vertical wind speed parameter *w* may be $< 0$ or $\geq 0$, whereas $\xi$ is always $\geq 0$. The gamma DSD prefactor parameter $\mu$ in *f(D)* may also be $< 0$ or $\geq 0$. Fig. 10a compares *f(D)* and *g(D)* for various values of the corresponding parameters for the negative case, or downdraft case. Figure 10b is the corresponding plot for the positive case, or updraft case. For $D \leq 0.1$ mm in Fig. 9a, $\mu = -0.3$ approximately corresponds to a downdraft velocity of $w = -1$ m s$^{-1}$, while $\mu = -0.5$ approximately corresponds to a downdraft velocity of $w = -3$ m s$^{-1}$. As *D* gets larger the similarity between *f(D)* and *g(D)* diverges quickly.

This is partly because for large $D$, $g(D) \to 1$, for $\xi = 0$ and for any $w$. Conversely for large $D$, $f(D) \to 0$ for $\mu < 0$ and $f(D) \to \infty$ for $\mu > 0$. Fig. 10b shows the effect of the parameter $\xi$, which is to populate the DSD with a non-zero fraction of the original exponential below the Heaviside cutoff. The importance of the regeneration parameter $\xi$ can be seen in this figure. When $\xi = 0$ and $w > 0$, the model predicts a complete absence of drop sizes $D$ below the range where the updraft velocity exceeds still air terminal velocity. Since disdrometer data seldom shows that kind of spectra, $\xi$ should be non-zero (and positive). This ad hoc parameter provides a simple solution to the problem. In future work, this part of the model should be replaced by something that better resembles the physics of the drop breakup mechanisms.

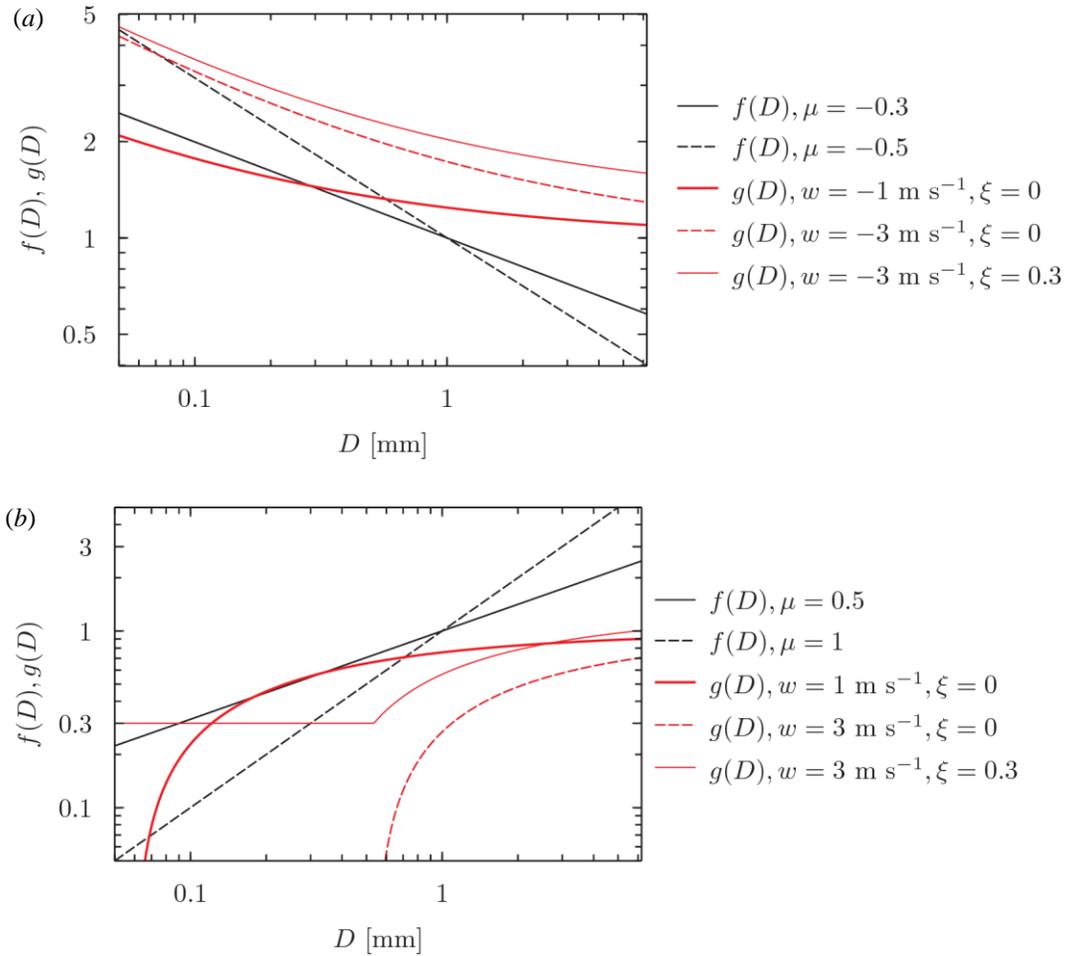

**Fig. 10**. Comparison of gamma DSD prefactor (dotted lines) and flux conservation equation prefactor, Eq. (21): (a) negative values of $\mu$ and $w$ (downdraft); (b) positive values of $\mu$ and $w$ (updraft).

Fig. 11a through 11c achieves this comparison directly by plotting the gamma DSD (Fig. 11a) and a plot generated from Eq. (18) (Fig. 11c) using parameters that produce similar behavior. Fig 11b plots a modified gamma distribution generated by defining a new $N_0' = N_0\, e^{-\mu H(\mu)}$, the goal being to realize a pseudo normalization for a better comparison to Eq. (18). Comparing Fig. 11b to 11c provides a comparison showing a qualitative relationship between the shape factor $\mu$ and vertical air motion $w$.

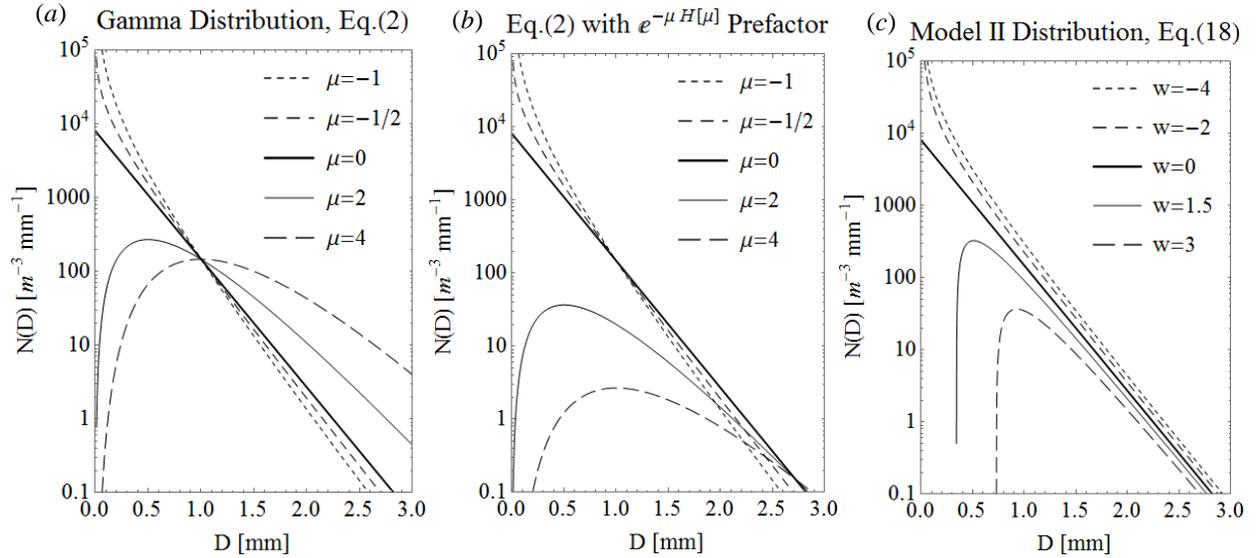

**Fig. 11**. Qualitative relationship between gamma distribution shape factor $\mu$ and vertical air motion $w$: (a) gamma DSD from Eq. (2); (b) modified gamma distribution using $N_0' = N_0\, e^{-\mu H(\mu)}$; (c) distribution from Eq. (18).

## 5. Summary

In this paper, a phenomenological relationship was investigated between vertical air motion and disdrometer derived $A$ and $b$ parameters corresponding to the well-known $Z$-$R$ power law. In this respect, three models were developed which are based on approximating raindrop terminal velocities using a modified version of the single particle trajectory modeling software, namely, Particle Trajectory with QShep (PTQ), developed at the NASA Granular Mechanics and Regolith Operations (GMRO) Lab, Kennedy Space Center, for the study of rocket-propelled regolith trajectories. The recently modified PTQ was used to simulate raindrop terminal velocities under conditions of still air and vertical winds. Disdrometer data from Athalassa, Cyprus was used to implement the different models developed in this study.

There appears to be a good correlation between the Model I simulations and the disdrometer data as shown in Figs. 6a and 6b. However, it must be recognized that the disdrometer data is separated into only two physical categories, stratiform and convective rain. We also know that there are three regions of disdrometer data in the $A$-$b$ domain:

stratiform, centered around $A = 300$ (grey symbols); convective rain where $A < 300$ (green symbols); and convective rain where $A > 300$ (red symbols). The model attempts to explain this separation by postulating that it is due to vertical wind, updraft or downdraft. Model I also takes a very simple approach to modifying the standard gamma DSD to account for vertical wind motion by discounting regeneration of drops and ignoring the surface boundary condition where the vertical wind component goes to zero. The later condition might be approximately true if the disdrometer were mounted at the top of a very tall tower, on the order of a 100 m or more. However, the data in Figs. 6a and 6b are from disdrometers located near the ground.

Model II takes into account more realistic conditions that correspond to surface conditions. This is a result of imposing flux conservation (or the continuity equation). In addition, a regeneration of drops below the Heaviside cutoff is accounted for by setting $\xi > 0$, but generally $\xi < 1$. Finally, Model II utilizes the results of the particle trajectory modeling which yields a nonzero $w$ component due to the time and distance required for the raindrop to adapt its terminal velocity to a change in vertical air speed as the moving air mass approaches the surface. Model II results are displayed in Fig. 7, analogous to the combination of Figs. 6a and 6b for Model I. In the simulation results (colored dots), a clear separation, but with substantial overlap, is seen due as a result of different $w$ values. Note that the uneven distribution of dots (Monte Carlo generated $A$-$b$ pairs) is mostly an artifact of the choice of parameter combinations ($\Lambda$, $N_0$, and $\xi$) in the Monte Carlo simulation algorithm. The disdrometer data, colored circles and squares, appears to have some correlation to the simulation.

Model III, as summarized by Eq. (24), is really just a sum of two Model II formulas. Instead of plotting the model extracted Monte Carlo $A$-$b$ pairs for various values of the model input parameters ($\Lambda_k$, $N_k$, and $\xi_k$ for $k = 1, 2$), fits were performed to the $N_D(D)$ disdrometer derived data that was acquired at the Athalassa site. Note that in order to compute $N_D(D)$ from the JW disdrometer histogram data, the raindrop terminal velocity formula of Eq. (10) was used with $w_0 = 0$. Fig. 9 shows the summarized results of Model III using the vertical velocities from the resulting model fits. The data in Fig. 9 partially follows the postulated trend that $A$-$b$ pairs are separated into three vertical regions along the $A$ axis attributed to the magnitude and sign of vertical air speed. However, there is also deviation from that trend, especially on the low $A$ side, for $w < 0$.

Table 1 shows $w > 0$ much more frequently than $w < 0$. However, data such as Kim and Lee (2016) show that same pattern when plotting the time series of vertical air motion retrieved from a 1290 MHz profiler-observed spectra. Their data examines in detail the passing of a stratiform rain system.

The techniques discussed thus far assume meaningful time averaged values of the DSD parameters corresponding to the entire rain event, 24 hours of rain events, or several months of rain events. This assumption runs into difficulties when the drop fall time $\tau$ from cloud level $h$ is on the order of the duration of the total acquired event $T$. In that case, it becomes necessary to process short time intervals such as one minute disdrometer data. This method is common when using the *method of moments* (MM) or other improved methods (Brawn and Upton, 2007). However, the MM fails when the disdrometer data is modeled as a sum of gamma distribution functions, which may be necessary in the long time average cases.

When considering short acquisition and processing times, Eq. (20) should be modified (letting $\xi = 0$, for simplicity) as follows:

$$N_D(D,t) = \chi\, H(\chi)\, N_0\bigl(t - \tau(D)\bigr) e^{-\Lambda(t-\tau(D))D} \quad , \tag{26}$$

where $\tau(D) = h/(v_T(D) - w)$. Eq. (26) requires considerable effort to process, but it might generate a much more detailed picture of the DSD. One approach for dealing with the difficulties of Eq. (26) is to process the disdrometer data using a convolution model approach, such as that described by Lane et al. (2009). The technique presented and discussed in this current work has thus far only been used to process time averaged data. The Cyprus data represents 24 h periods starting at midnight. No attempt was made to select individual rain events or worry about events that span midnight. The data published in Sulochana et al. (2016) represents seasonal averages, which involves several months of data for each *A-b* pair plotted. One minute intervals would show more detail and would likely not require the dual sum used to define Model III. However there is a problem with the flux conservation model and short sampling intervals and that is fall time as a function of drop size. A possible area for future work follows by using Eq. (26) on short time data, such as one minute intervals.

Another strategy for future work is to collocate a vertical wind measurement system, such as an acoustic SODAR, with the disdrometer. Then, actual values of *w* can be used in the analysis. Fig. 9 tends to provide some confidence that *w* affects *A-b* in a partially predictable way. Fig. 9 also strongly suggest that additional mechanisms are significant and at work. These mechanisms should be investigated and accounted for in the model, in addition to the effect of vertical wind motion. Other potential mechanisms might include, drop coalescence, evaporation, and *superterminal* drops (see Montero-Martínez et al., 2009).

The flux conservation model, described by Eq. (17), is an idealized concept. Since hydrometeor flux is the product of a size distribution $N(D)$ and a velocity function $v_D(D)$, the notion that it is conserved requires that all hydrometeors in $N(D)$ remain intact and undisturbed, except for velocity changes. This constraint is generally unrealistic due to evaporation and collisions, resulting in breakup and coalescence. However, the overall effect of flux conservation may be observed when a balance between breakup and coalescence occurs, i.e., when the DSD is in an equilibrium state. Flux will also be conserved when the surfaces that are being compared are separated by small distances. A flux plane at cloud level compared to a surface on the ground is the worst case. For future work, it may be more useful to engage the differential form of flux conservation:

$$\mathbf{v}_D \nabla N_D = -N_D \nabla \cdot \mathbf{v}_D \quad , \qquad (27)$$

where $\mathbf{v}_D$ and $N_D$ are in general a functions of $x$, $y$, $z$, and $D$.

The flux conservation concept implies that the still air observed DSD is composed of two parts, a zero motion DSD and a drop velocity factor. The product of the two is the drop flux distribution (DFD) and is a pseudo conserved quantity. The observed DSD (by radar or optical means) stretches and compresses based on vertical wind motion and still air drop terminal velocity. A thought experiment for this is to imagine a 1D case: a line of billiard balls of diameter $D_0$, each separated by a distance $a$, rolling down an incline at constant velocity or *feed rate* $\eta$, then free falling after exiting the end of the incline (high above the ground). The initial 1D DSD is $N_i(D) = a^{-1} \delta(D-D_0)$, where $\delta(D-D_0)$ is the Dirac delta function. After the balls have been in free fall and obtained terminal velocity $v_T$, the observed 1D DSD (observed by radar or optical means) is $N(D) = (\eta/v_T) a^{-1} \delta(D-D_0)$. The terminal velocity term in the denominator leads to the negative shape factor in a gamma distribution representation.

In granular mechanics, a normalized size fraction of a pile of soil can be represented by $S(D)$ with units of mm$^{-1}$. When $S(D)$ is normalized, its integral from 0 to ∞ over $D$ is equal to 1 (see Eq. (2) in Lane and Metzger, 2015). Now imagine a large reservoir of soil with a port at the bottom, high above the ground (it could be at cloud level). All particles are passing through the port at some feed velocity $\eta$. After some time all particles are in free fall and have reached their terminal velocities, Considering perfectly still air, the product of velocity $v(D)$ (equal to particle terminal velocity in still air) and the aerial size distribution $N(D)$ divided by the integral of $v(D)$ $N(D)$ from 0 to ∞ over $D$, is equal to $S(D)$ at any time during equilibrium conditions (non-equilibrium conditions exists at the start and stop of the event). If the size of the reservoir is infinite, then after the smallest particles have made their way to the measurement surface (e.g. ground), the flux is conserved from that time forward and the measured flux divided by its integral is equal to $S(D)$.

There are two important points in the above discussion. First, flux conservation is conserved, but only under certain conditions. Therefore flux conservation for hydrometeors is at best a pseudo conservation law. The second point is that even though the concept of a pile of hydrometeors is non-sensible, the implications of the idea may be useful. For lunar soil or any other granular system, the size distribution $S(D)$ is a fundamental quantity, whereas $N(D)$ is not. $N(D)$ for blowing lunar dust is dependent on both $S(D)$ and $v(D)$. It may be useful then to treat hydrometeors at the generation source (within a cloud) as a pile of hydrometeors, described by $S_H(D) = \Lambda\, e^{-\Lambda D}$. It follows that $N(D) = n\, S_H(D)/v(D)$, where $n$ is the total flux defined by the integral of $v(D)\, N(D)$ over all $D$. If $v(D)$ is represented by a simple power law, $v(D) = a\, D^b$, and $N_0$ is defined as $n/a$, the result is $N(D) = D^{-b} N_0\, e^{-\Lambda D}$, the familiar gamma distribution with a negative shape factor.

## 6. Conclusion

The goal of this work was to investigate the distribution of disdrometer derived points in the *A-b* parameter plane. Eqs. (5) and (6) express precise values of *A* and *b* when the DSD is described by an exponential distribution and the drop velocity function is described by a power law. The *A* expression of Eq. (6) is dependent on the $N_0$ parameter of the DSD. Therefore, the cluster of *A-b* points represents a horizontal line in the *A-b* plane under these ideal conditions, where $N_0$ is allowed to vary. The well-known gamma distribution of Eq. (2) generates a continuous set of points in the *A-b* plane as shown in Fig. 5a by varying the DSD parameter values. In both the exponential and gamma DSD cases, *A* is independent of the parameter $\Lambda$ only when the limits of integration for *R* and *Z* are 0 to $\infty$. When the limits of integration are based on realistic drop size limits, such as $D_{min}$ = 0.3 mm and $D_{max}$ = 5.5 mm, the parameter *A* is coupled to $\Lambda$.

After this baseline was established, the next goal was to examine the effect of vertical air motion and the subsequent effect on drop terminal velocity, knowing that many other factors are contributors, but ignoring their influence for simplicity. Using a trajectory model of particle motion under the influence of gravity in a fluid which may also be moving, it was shown that vertical air motion above the ground can have an effect on the terminal velocity of particles at the surface. The effect of vertical air motion on particle velocity is greatly reduced at the surface, but has the potential of having an observable influence on a hydrometeor's final velocity. The effect of vertical air motion is to increase (for an downdraft) or decrease (for an updraft) the drop velocity $v_D(D)$, which is equal to the still air terminal velocity $v_T(D)$ as defined for zero vertical air motion.

A separate part of this investigation included the concept of a raindrop flux conservation model and its influence on the drop size distribution. A key point of this model is that terminal velocity and drop size distribution are independent mechanisms, but are very much coupled through this pseudo conservation law. For example, the drop size distribution is the result of a DSD generation rate within a cloud, but the DSD observed aloft, beneath the cloud and away from the rain generator's influence, is modified by the drop velocity function. The particle generation function and the particle terminal velocity function are independently the result of the physical processes of the planet's atmosphere and gravity constant.